\documentclass[aps,prl,amsmath,amssymb, reprint,superscriptaddress,nofootinbib]{revtex4-2}

\usepackage{graphicx}
\usepackage{dcolumn}
\AtBeginDocument{\usepackage{booktabs}}
\usepackage{units}
\usepackage{hyperref}
\usepackage{soul}
\usepackage{braket}
\usepackage{color}
\definecolor{red}{RGB}{255,0,0}

\usepackage{comment}
\usepackage{enumitem}
\begin{document}
	
\preprint{}

\newcommand{\thetitle}{Hybrid entanglement and error correction in a scalable quantum network node}
% possible titles
% Hybrid entanglement in a solid-state quantum register
% Hybrid quantum entanglement (and error correction) in a solid-state quantum network node
% Hybrid quantum entanglement (and error correction) in a solid-state quantum register

\author{Xiu-Ying Chang}
\thanks{These authors contributed equally to this work.}
%\altaffiliation{These authors contributed equally to this work.}
\affiliation{Center for Quantum Information, Institute for Interdisciplinary Information Sciences, Tsinghua University, Beijing 100084, People’s Republic of China}
\affiliation{Hefei National Laboratory, Hefei 230088, PR China}

\author{Pan-Yu Hou }
\email[]{houpanyu@tsinghua.edu.cn}
%\thanks{These authors contributed equally to this work.}
\affiliation{Center for Quantum Information, Institute for Interdisciplinary Information Sciences, Tsinghua University, Beijing 100084, People’s Republic of China}
\affiliation{Hefei National Laboratory, Hefei 230088, PR China}

\author{$^{,\,*}$ Wen-Gang Zhang}
\thanks{These authors contributed equally to this work.}
\affiliation{Center for Quantum Information, Institute for Interdisciplinary Information Sciences, Tsinghua University, Beijing 100084, People’s Republic of China}
\affiliation{Beijing Academy of Quantum Information Sciences, Beijing 100193, China}

\author{Xiang-Qian Meng} 
\affiliation{Center for Quantum Information, Institute for Interdisciplinary Information Sciences, Tsinghua University, Beijing 100084, People’s Republic of China}
\affiliation{Hefei National Laboratory, Hefei 230088, PR China}

\author{Ye-Fei Yu}
\affiliation{Center for Quantum Information, Institute for Interdisciplinary Information Sciences, Tsinghua University, Beijing 100084, People’s Republic of China}
\affiliation{Beijing Academy of Quantum Information Sciences, Beijing 100193, China}

\author{Ya-Nan Lu}
\affiliation{Center for Quantum Information, Institute for Interdisciplinary Information Sciences, Tsinghua University, Beijing 100084, People’s Republic of China}
\affiliation{Hefei National Laboratory, Hefei 230088, PR China}

\author{Yan-Qing Liu}
\affiliation{Center for Quantum Information, Institute for Interdisciplinary Information Sciences, Tsinghua University, Beijing 100084, People’s Republic of China}
\affiliation{Hefei National Laboratory, Hefei 230088, PR China}

\author{Bin-Xiang Qi}
\affiliation{Center for Quantum Information, Institute for Interdisciplinary Information Sciences, Tsinghua University, Beijing 100084, People’s Republic of China}
\affiliation{Hefei National Laboratory, Hefei 230088, PR China}

\author{Dong-Ling Deng}
\email[]{dldeng@tsinghua.edu.cn}
\affiliation{Center for Quantum Information, Institute for Interdisciplinary Information Sciences, Tsinghua University, Beijing 100084, People’s Republic of China}
\affiliation{Hefei National Laboratory, Hefei 230088, PR China}
\affiliation{Shanghai Qi Zhi Institute, No. 701 Yunjin Road, Xuhui District, Shanghai 200232, China}

\author{Lu-Ming Duan}
\email[]{lmduan@tsinghua.edu.cn}
\affiliation{Center for Quantum Information, Institute for Interdisciplinary Information Sciences, Tsinghua University, Beijing 100084, People’s Republic of China}
\affiliation{Hefei National Laboratory, Hefei 230088, PR China}
\affiliation{Shanghai Qi Zhi Institute, No. 701 Yunjin Road, Xuhui District, Shanghai 200232, China}
\affiliation{New Cornerstone Science Laboratory, Beijing 100084, PR China}

\title{\thetitle}
\date{\today}

\begin{abstract}
    Recent breakthroughs have ushered the quantum network into a new era, where quantum information can be stored, transferred, and processed across multiple nodes on a metropolitan scale. 
    A key challenge in this new era is enhancing the capabilities of individual nodes, providing precise and robust control over multiple qubits and advanced functionality for scalable quantum networks. 
    Here, we report on precise and complex control in a hybrid quantum node based on a diamond color center. We demonstrate hybrid coherent control by entangling three types of qubits: an electron spin as an interface qubit, a nuclear spin with long memory time, and a flying photonic qubit, with their qubit frequencies spanning three distinct regimes from the optical domain to the rf domain. 
    By incorporating two additional memory qubits, we encode three memory qubits into a logical state using the three-qubit repetition code and entangle this logical qubit with a photonic qubit.  
    Leveraging hybrid qubits and precise control, we repeatedly read out the error syndromes of memory qubits through the electron spin, serving as an auxiliary qubit, then apply a real-time feedback operation to correct bit-flip errors. 
    We execute and verify active error correction for up to twelve rounds and demonstrate the improvement over the uncorrected counterpart. Our results demonstrate the feasibility of several key functionalities for next-generation quantum repeaters, paving the way towards full-fledged metropolitan-scale quantum networks for a wide range of practical applications.
    %
    %Our results demonstrate several key features of next-generation quantum repeaters, potentially facilitating applications of metropolitan-scale quantum networks. The approach can be further enhanced by improving operation fidelity and by including more qubits for advanced error correction codes. It can be adapted to other quantum network platforms, such as trapped ions and neutral atoms.
\end{abstract}
\maketitle

\section{Introduction}

% Background
Future quantum network will process quantum information across a large number of quantum registers linked through shared quantum entanglement, holding the promise to revolutionize a broad range of technologies, such as quantum key distribution~\cite{ekert1991quantum,lo2014secure}, distributed quantum computing~\cite{buhrman2003distributed,kimble2008quantum}, non-local metrology~\cite{komar2014quantum, nichol2022elementary}, among others~\cite{kimble2008quantum, wehner2018quantum, azuma2023quantum}. 
% previous efforts
In recent decades, increasing endeavors have been made to develop prototypes of quantum network nodes in various platforms including trapped ions~\cite{moehring2007entanglement,duan2010colloquium} and atoms~\cite{chou2005measurement, chou2007functional, yuan2008experimental,hofmann2012heralded}, diamond color centers~\cite{bernien2013heralded,childress2013diamond,ruf2021quantum,sipahigil2016integrated}, and quantum dots~\cite{delteil2016generation,stockill2017phase,lodahl2017quantum}.
Elementary functionalities of individual nodes have been demonstrated for establishing remote entanglement~\cite{bernien2013heralded,sipahigil2016integrated,stephenson2020high,yu2020entanglement} as well as local information processing and storage~\cite{stas2022robust,inlek2017multispecies}. Several key primitives have been realized in laboratory-scale quantum networks~\cite{kalb2017entanglement, humphreys2018deterministic, jing2019entanglement,  pompili2021realization, hermans2022qubit,bhaskar2020experimental}. 
These advancements have led to recent breakthroughs, the realizations of metropolitan-scale quantum networks~\cite{liu2024creation,knaut2024entanglement,stolk2024metropolitan}, illuminating the potential for practical applications in the near future. 

\begin{figure*}[tp!]
\includegraphics[width=1.0\linewidth]{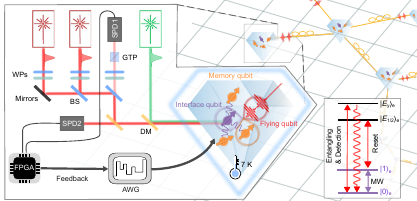}
\caption{A hybrid multi-qubit node in a large-scale quantum network. In a diamond-based quantum network (right side), different nodes are connected through optical fiber links (yellow lines). Each node contains a hybrid quantum register associated with an NV center. The electron spin, serving as an interface qubit (purple), can be entangled with single photons (red wavy lines) and interacts with nearby long-lived nuclear spins (orange). The relevant energy level structure and control fields are illustrated in the inset.
Left grey box shows a simplified experimental setup. A 532\,nm laser (green) reset the NV charge state. Two red lasers near 637\,nm drive are used for entanglement generation, state-dependent fluorescence detection, and electron spin state reset. 
All laser beams are combined and focused by an objective onto a diamond sample located in a cryogenic environment at approximately 7\,K. NV fluorescence is collected by the same objective and detected by two single-photon detectors (SPDs), one for ZPL photons and the other for phonon-sideband photons. A fast FPGA board determines feedback operations in real time by controlling the AWG output for error correction and perform spin-photon joint measurements. 
WPs: waveplates; BS: beamsplitter; DM: dichroic mirror; GTP: Glan-Taylor prism; AWG: arbitrary wave generator; MW: microwave. 
}\label{fig:fig1}
\end{figure*}

Practical large-scale quantum networks, which connect multiple locations through optical fiber infrastructure (illustrated in Fig.~\ref{fig:fig1}) will likely rely on quantum repeater protocols~\cite{briegel1998quantum,duan2001long,sangouard2011quantum}. This requires many network nodes, each of which must fulfill the following requirements to handle the increasing difficulty as the network scales up~\cite{wehner2018quantum, ruf2021quantum,azuma2023quantum}. 
First, each node should contain at least one interface qubit, capable of entangling with photons to establish remote entanglement. 
Second, each node must include memory qubits with long coherence times to store quantum states during the time needed to generate entanglement across the entire network for the full network activity, not just between neighboring nodes.
Third, each node should have multiple memory qubits to support advanced functionalities for maintaining the integrity of quantum information, such as error correction. These functionalities are crucial as the network scales up. An increased number of nodes and time for entire network activity will result in larger errors due to imperfect operations and decoherence, inevitably degrading network performance. Multiple memory qubits also facilitate multiplexing, enhancing the efficiency of entanglement generation.
Furthermore, using different types of qubits as interface and memory qubits offer advantages for robust performance, as operations on one type are less likely to cause cross-talk errors on the other. However, controlling and measuring different types of qubits is highly nontrivial and challenging due to their distinct physical properties and experimental control complexity. 
These capabilities have been individually demonstrated on platforms such as diamond color centers~\cite{childress2013diamond,waldherr2014quantum,ruf2021quantum,cramer2016repeated,abobeih2022fault} and trapped atoms~\cite{moehring2007entanglement,duan2010colloquium,ryan2021realization}. However, integrating all these features into a single quantum network node remains a substantial challenge. 

% Key points of what we did. 
Here we present a hybrid multi-qubit quantum network node based on a nitrogen-vacancy (NV) center in diamond,  implementing several key functionalities required for scaling up quantum networks. 
Our main results are two-fold. 
First, we demonstrate quantum control of this hybrid node by generating a Greenberger–Horne–Zeilinger-type (GHZ) state involving three different types of qubits---an electron spin  serving as an interface qubit, a carbon-13 nuclear spin as a memory qubit, and a photon as a flying qubit---with fidelity exceeding 0.83(3). 
This entangled state connects qubits across very distinct frequencies in the rf, microwave, and optical domains, making it a crucial resource for network applications. 
Second, by harnessing hybrid control and incorporating additional memory qubits, we encode the memory qubits in a logical state and entangle it with a flying qubit. In this proof-of-principle experiment, we use a three-qubit repetition code to protect memory qubits from bit-flip errors. This is accomplished by repeatedly reading out error syndromes through the interface qubit and applying active feedback to correct errors. We verify error correction by comparing the final states with and without active feedback. Furthermore, we show that the final states can be improved by post-selecting results where the error syndrome outcomes indicate no errors, i.e. error detection.  
These results showcase the key advance of hybrid nodes in diamond and their potential for realizing full-fledged metropolitan-scale quantum networks.

\section{Experimental setup}
 
Our hybrid quantum network node is based on a single NV center in a high-purity synthetic diamond located in a cryogenic environment. The relevant level structure and a simplified experimental setup are shown in Fig.~\ref{fig:fig1} (see details in Appendix).
% interface qubit: electron spin 
The interface qubit is encoded in two sublevels of the NV electron spin, $\ket{m_s=0} \equiv \ket{0}_e$ and $\ket{m_s={-1}} \equiv \ket{1}_e$, which can be coherently manipulated by resonant microwave fields.  State initialization, spin-photon entanglement, and state-dependent fluorescence readout are achieved through resonant optical excitation. 
Single-shot readout fidelity is asymmetric, 0.809(3) for the ``bright" state $\ket{0}_e$ and 0.988(1) for the ``dark" state $\ket{1}_e$. Coherent photons at the zero-phonon line (ZPL) and state-dependent fluorescence are detected by two separate detectors~\cite{robledo2011high}, with signals recorded by an field-programmable gate array (FPGA) board for time-resolved detection and correlation measurements. 
Carbon-13 nuclear spins near the NV center, which have long coherence times, are used as memory qubits: $\ket{m_I=-1/2} \equiv\ket{0}_c$ and  $\ket{m_I=1/2} \equiv\ket{1}_c$. 
Universal control of individual nuclear spins is accomplished by manipulating the electron spin with tailoring microwave pulses~\cite{taminiau2014universal}. State initialization and readout of nuclear spins are implemented by swapping their states with the electron spin. 

\begin{figure*}
\includegraphics[width=1.0\textwidth]{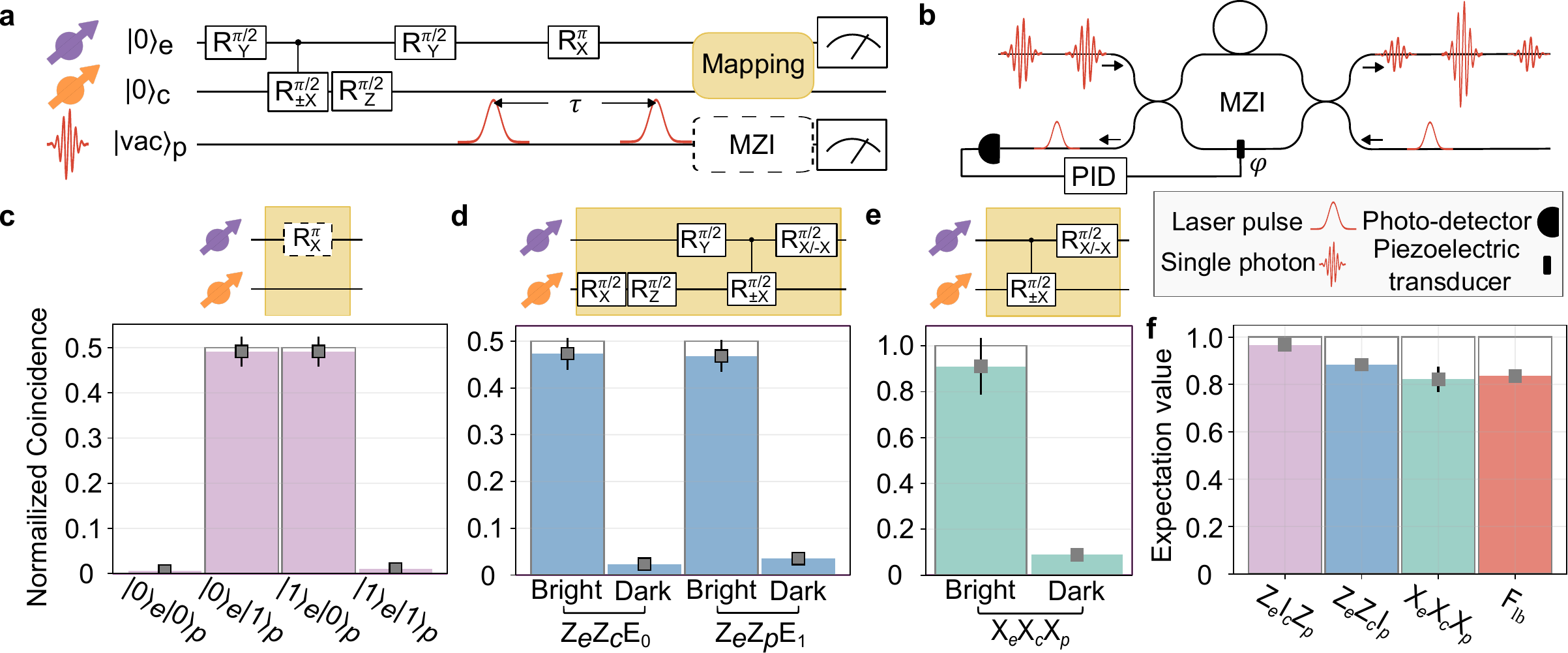}
\caption{Generation and characterization of a hybrid GHZ state. \textbf{a}, Experimental pulse sequence. Electron and nuclear spins are initialized in $\ket{0}_e\ket{0}_c$ then entangled with single-qubit rotations and entangling gates. Two optical $\pi$ pulses with a fixed separation of $\tau\approx$~52\,ns and a microwave $\pi$ pulse in between entangle the local spins with a time-bin photonic qubit. Three-qubit correlations are measured in three different bases.  
The information about the electron-nuclear spin states is mapped 
onto the electron spin by the sequences shown above the bar plots in \textbf{c}-\textbf{e}, then read out by fluorescence detection. 
Photonic qubit states $\ket{0}_p$ and $\ket{1}_p$ are identified by time-resolved single-photon detection. For phase measurement, the single photon is sent through an MZI.  
Coincidence events of two detections are recorded to extract three-qubit correlations. 
\textbf{b} An imbalanced MZI is used to access phase coherence of the time-bin qubit. Early and late time-bin components can partially overlap and interfere at the MZI outport (the middle wavy line at the upper right port). The arm length difference is stabilized by monitoring the laser intensity of a red laser traveling through the MZI and applying a feedback voltage from a servo controller to a piezo transducer on one arm. 
\textbf{c}-\textbf{e} Normalized coincidence counts are shown as bars along with 68\%-confidence error bars for the measurements of (\textbf{c}) $\rm Z_eI_cZ_p$, (\textbf{d}) $\rm Z_eZ_cI_p$, (\textbf{e}) $\rm X_eX_cX_p$. Even parity of $\rm Z_{e} Z_{c}$ (\textbf{d}) and $\rm X_{e} X_{c}$ (\textbf{e}) is mapped onto the bright $\ket{0}_e$ and dark $\ket{-1}_e$ states respectively. In \textbf{d}, $\rm M_{p}^{e}=\ket{e}_{p}\bra{e}$, $\rm M^{l}_{p}=\ket{l}_{p}\bra{l}$. 
\textbf{f} Expectation values of the three measurement bases and the lower bound of the GHZ state fidelity are estimated using the data in (\textbf{c-e}).}\label{fig:fig2}
\end{figure*}

\section{Hybrid entanglement generation}
To demonstrate multi-qubit quantum control in the hybrid network node, we prepare three different types of qubits in a GHZ-type state and verify genuine multipartite entanglement using the pulse sequence shown in Fig.~\ref{fig:fig2}\textbf{a}. 
% Entangle electron spin with nuclear spin
The electron-nuclear spin register is initialized in $\ket{0}_e \ket{0}_c$. 
A microwave $\pi/2$ pulse alters the electron spin to a superposition state $\frac{1}{\sqrt{2}}\left(\ket{0}_e+\ket{1}_e\right)$ while leaving the nuclear spin unchanged. 
Subsequently, an entangling gate $R_{\rm \pm  X}^{\pi/2}$ rotates the nuclear spin around the $\rm +X$ axis for an angle of $\pi/2$ on the Bloch sphere when the electron spin is in $\ket{0}_e$ while around the opposite axis $\rm -X$ for the same angle when in $\ket{1}_e$. Then, two single-qubit rotations respectively change the nuclear and electron spin states, resulting in the desired Bell state $\frac{1}{\sqrt{2}}\left(\ket{0}_e\ket{1}_c+ \ket{1}_e\ket{0}_c\right)$. 
% Entangle local spins with photon
We further entangle the local spin register with a single photon encoded in time bins~\cite{bernien2013heralded}. 
A resonant optical $\pi$ pulse selectively drives the $\ket{0}_e \ket{0}_c$ state to the excited state $\ket{E_y}_e \ket{0}_c$, accompanied by subsequent single-photon emission. %The nuclear spin state is unchanged with a high probability during this process. 
We then flip the electron spin using a microwave $\pi$ pulse and repeat the optical pumping process. The second component in the Bell state is correlated with a single photon emitted at a later time. 
A GHZ state in the form of $\ket{\phi}_{\rm GHZ}=\frac{1}{\sqrt{2}} (\ket{0}_e\ket{0}_c\ket{l}_p)+\ket{1}_e \ket{1}_c \ket{e}_p)$ is generated, where the basis states $\ket{e}_p$ and $\ket{l}_p$ correspond to a single-photon emission event at an earlier or later time, separated by approximately 52\,ns.

% GHZ state characterization
Characterizing multiple qubits of different types requires the detection of their correlations.  
We address this challenge by performing joint-state measurements in three different bases $\rm Z_{\it e}I_{\it c}Z_{\it p}$, $\rm Z_{\it e} Z_{\it c} I_{\it p}$, and $\rm X_{\it e} X_{\it c} X_{\it p}$, where Pauli operators $\rm X$, $\rm Z$ and identity operator $\rm I$ are with subscripts denoting the corresponding qubits. Specifically, $\rm Z_{\it p} = \it\ket{E}_{\it p}\bra{E} - \ket{L}_{\it p}\bra{L}$.
As the generated GHZ state is an eigenstate of these operators, a lower bound of the fidelity to the ideal state can be derived from the stabilizer witness method and calculated by~\cite{somma2006lower,toth2005entanglement,guhne2009entanglement}
\begin{equation}\label{eq:eq2}
F_{lb}=\frac{1}{2} \left[ \langle -\rm Z_{\it e}I_{\it c}Z_{\it p} \rangle + \langle \rm Z_{\it e} Z_{\it c} I_{\it p} \rangle + \langle X_{\it e} X_{\it c} X_{\it p} \rangle - 1\right] 
\end{equation} 

% how we do these measurements
In our experiments, the joint measurements are accomplished by correlating the outcomes of two detections: one for the local spins and the other for the single photon. The former involves a microwave pulse sequence that maps the information about certain electron-nuclear parity onto the electron spin population followed by state-dependent fluorescence detection (see the mapping sequences in Figs.~\ref{fig:fig2}\textbf{c}-\textbf{e}). 
Because this fluorescence detection has asymmetric readout fidelity~\cite{robledo2011high}, for each measurement basis we choose to perform two sets of experiments where the electron spin is either untouched or flipped after the mapping sequence and before fluorescence detection so that both electron spin states at the end of the mapping can be bright during the detection.  
Readout of the photonic time-bin qubit in the $\rm Z$ basis is achieved using time-resolved detection enabled by a fast FPGA board, which also records the coincidence events where both detections register at least one photon. Expectation values of the three observable are calculated from the measured coincidence counts.
Details of the joint measurements can be found in Appendix.

% Results of GHZ
The results of the GHZ state characterization are shown in Figs.~\ref{fig:fig2}\textbf{c}-\textbf{f}. 
% ZIZ 
For $\rm \langle Z_{\it e}I_{\it c}Z_{\it p} \rangle$, we obtain the normalized coincidence counts of the four electron-photon states, which are shown in Fig.~\ref{fig:fig2}\textbf{c}.  
The sum population of the target states, $\ket{0}_e\ket{l}_p$ and $\ket{1}_e\ket{e}_p$, is approximately 0.98 with an error of 0.02 due to the background and the finite separation between time bins, yielding $\langle \rm Z_{e}I_{c}Z_{p} \rangle=0.97(1)$. 
%  ZZI 
Although the photonic qubit state is irrelevant for measuring $\langle\rm Z_{e}Z_{c}I_{p}\rangle$, the photon presence must be verified. Therefore, we measure the electron-nuclear parity of $\rm Z_{e}Z_{c}$ conditioned on successful single-photon detection at either time bin. As shown in Fig.~\ref{fig:fig2}\textbf{d}, similar results are observed conditioning on both time-bin photon detection, as expected. The experimental $\langle \rm Z_{e}Z_{c}I_{p}\rangle$ is estimated to be 0.88(2) using the normalized coincidence counts, with errors dominated by nuclear spin operation errors.  
% XXX
Measuring $\langle \rm X_{e}X_{c}X_{p} \rangle$ involves assessing the phase coherence of the time-bin qubit. This is done by sending the single photon through an imbalanced Mach-Zehnder interferometer (MZI, shown in Fig.~\ref{fig:fig2}\textbf{b}), allowing the early and later time-bin components to overlap and interfere at the MZI output ports. The arm length difference of the MZI is stabilized between experiments (see Appendix). 
We estimate $\langle \rm X_{e}X_{c}X_{p} \rangle$ to be 0.82(5). Besides the error sources due to dark counts and nuclear spin control, additional errors in this measurement arise from fluctuations and drifts in the MZI. 
% GHZ fidelity
Finally, we compute the lower bound of the fidelity to an ideal GHZ state using Eq.~(\ref{eq:eq2}), yielding $F_{lb}=0.84(3)$ without subtracting qubit initialization and measurement errors. This value significantly surpasses the classical bound of 0.5, indicating genuine multipartite entanglement among three distinct degrees of freedom.
 
\section{Repeated error correction} 

\begin{figure*}
\includegraphics[width=1.0\textwidth]{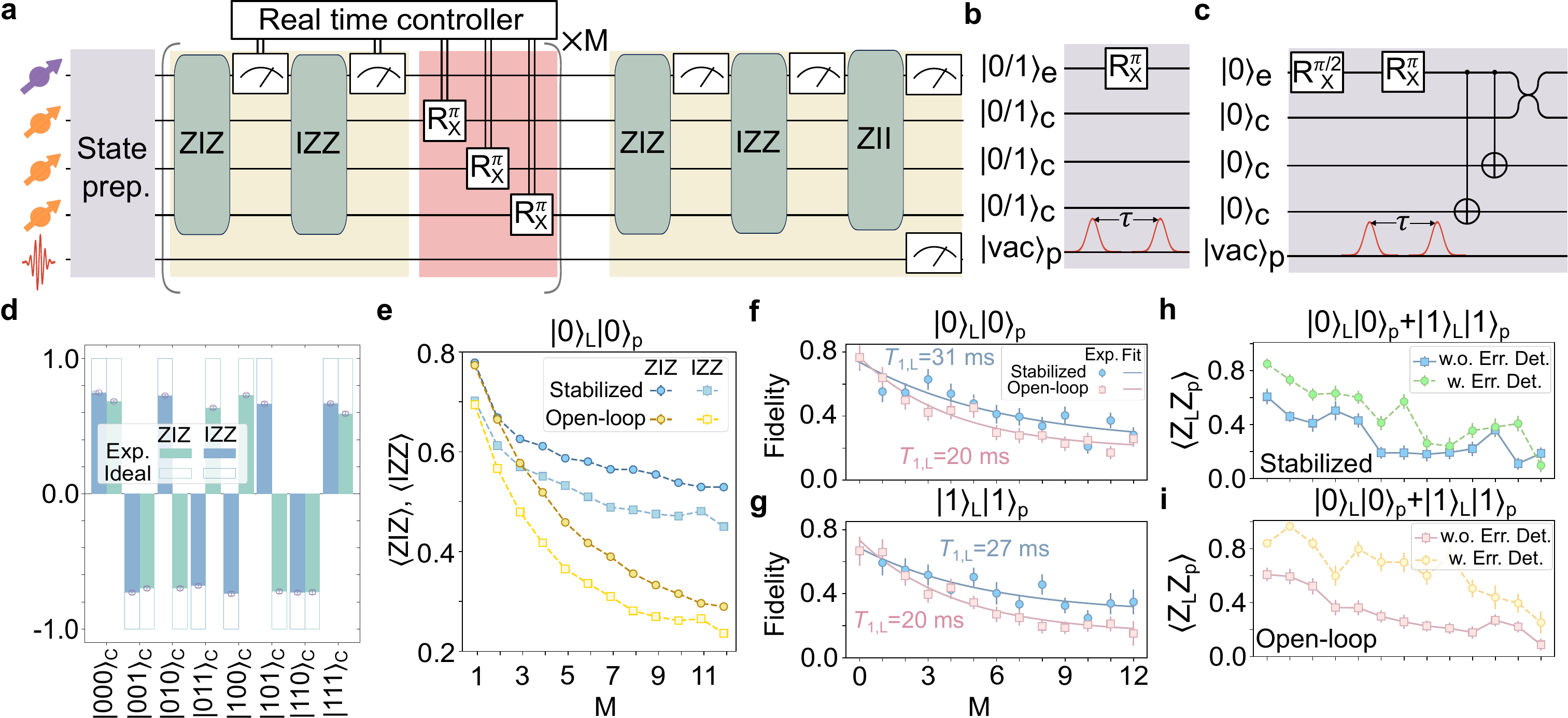}
\caption{Repeated error correction and error detection of local memory qubits in a hybrid network node. \textbf{a}, Experimental pulse sequence. A logical qubit encoded in three nuclear spins (orange spins) is entangled with a photonic qubit (red wavy line) using the sequence shown in \textbf{b}. Single bit-flip error syndromes of the local memory qubits, parity checks of $\rm ZIZ$ and $\rm IZZ$, are repeatedly read out through the electron spin (purple spin). The syndrome measurement outcomes are sent to a real-time controller that determines the feedback operation to correct the error. The final state is projected onto a product state $\ket{i_1,i_2,i_3}_c\ket{j}_p$ , with $i_1, i_2, i_3 \in\{0,1\}$ and $j \in \{e,l\}$, by three sequential measurements of nuclear spins in bases $\rm ZIZ$, $\rm IZZ$, $\rm ZII$, all through the electron spin, and the photonic qubit measurement in the $\rm Z$ basis. As $\rm ZII=Z_{L}$, $\langle \rm Z_{L}Z_{p}\rangle$ of the final state can be extracted with error detection using the outcomes of the parity checks in the final measurement. 
\textbf{b}, Pulse sequence for generating an entangled logical-photon state $(\ket{0}_L\ket{e}_{p} + \ket{1}_L\ket{l}_{p})/\sqrt{2}$. 
\textbf{c}, Pulse sequence for generating product states  $\ket{0}_{L}\ket{e}_p$ and $\ket{1}_{L}\ket{l}_p$. 
\textbf{d}, Characterization of the parity checks. Experimental values (solid bars) with fluorescence detection error excluded are compared to the ideal values (empty bars). 
\textbf{e} Average values of the parity checks with (blue symbols, labeled as ''stabilized") and without (orange symbols, ``open-loop") feedback are plotted against the error correction round index $\rm M$. 
\textbf{f}, \textbf{g}, Fidelity of the final state to an initial state of $\ket{0}_{L}\ket{e}_p$ (\textbf{f}) or $\ket{1}_{L}\ket{l}_p$ (\textbf{g}) versus $\rm M$, with (blue dots) and without (pink squares) feedback, are plotted along with the fits to an exponential decay function. 
\textbf{h,i} With an initial state $(\ket{0}_L\ket{e}_{p} + \ket{1}_L\ket{l}_{p})/\sqrt{2}$, the average values of the logical operator $\langle \rm Z_{L}Z_{p}\rangle$ with (dots) and without (squares) error detection are extracted from the final measurement. Both stabilized (h) and open-loop (i) cases show higher $\langle \rm Z_{L}Z_{p}\rangle$ using error detection. The improvement is more pronounced in the open-loop case. 
%.
}\label{fig:fig3}
\end{figure*}

% Dive into QEC results. 
Multiple long-lived nuclear spins can be found near an NV center, providing essential qubit resources for quantum error correction. As a proof-of-principle experiment, we encode a logical qubit using three carbon-13 nuclear spins that weakly couple to the NV electron spin. 
% Introduce repetition code and its basics.
More concretely, we employ the three-qubit repetition code~\cite{terhal2015quantum}, where the logical states are defined as $\ket{0}_{L}\equiv \ket{000}_{c}$  and $\ket{1}_{L}\equiv \ket{111}_{c}$.  This code is designed to correct single bit-flip errors by using a majority vote of the three qubit values and flipping the qubit that differs from the majority. The majority vote is implemented through measurements of two parity checks, specifically, $\rm ZIZ$ and $\rm IZZ$. If $\rm ZIZ=-1$ and $\rm IZZ=+1$, it indicates a single-bit error occurs on the first qubit and can be corrected by flipping this qubit. If $\rm ZIZ=+1$ and $\rm IZZ=-1$, the error is on the second qubit.  If both equal $-1$, the error is on the third. Both equal $+1$ indicates no error. Note that this code cannot correct multiple errors or phase-flip errors. We focus on demonstrating the key components of general error correction protocols for network applications and potential improvement in local spin relaxation. 

The experimental pulse sequence for repeated error correction of local qubits is depicted in Fig.~\ref{fig:fig3}\textbf{a}. 
% State preparation 
For quantum network applications, encoded logical qubits must be entangled with photonic qubits to enable remote entanglement between error-correctable logical qubits. 
As shown in Fig.~\ref{fig:fig3}\textbf{b}, we address this challenge through the following steps: 1) generating an electron-photon entangled state $\alpha \ket{0}_e\ket{e}_p+ \beta\ket{1}_e\ket{l}_p$, 2) applying two CNOT gates to entangle them with two memory qubits, resulting in the state $\alpha \ket{0}_e\ket{e}_p\ket{00}_c+ \beta\ket{1}_e\ket{l}_p\ket{11}_c$, 3) coherently swapping states between the electron spin and the third memory qubit, yielding $\ket{0}_e(\alpha\ket{e}_p\ket{000}_c+ \beta\ket{l}_p\ket{111}_c)=\ket{0}_e(\alpha\ket{e}_p\ket{0}_L+ \beta\ket{l}_p\ket{1}_L)$ with the electron spin disentangled from the rest. The generated logical-photon entangled state is a $+1$ eigenstate of $\rm Z_{L} Z_{p}$, where $\rm Z_{L}$ is a logical-$\rm Z$ operator. The CNOT gates and swap operations are composed of multiple elementary gates, with details and characterization results provided in Appendix. Product states $\ket{0}_L\ket{e}_{p}$ and $\ket{1}_L\ket{l}_{p}$ are prepared in a simpler way for higher fidelity using the sequence shown in Fig.~\ref{fig:fig3}\textbf{c}.  
% Stabilizer measurement characterization
The two parity checks (green blocks) are sequentially read out through the electron spin using a designated mapping sequence followed by state-dependent fluorescence detection. For both parity checks, we choose to map the eigenvalue of $+1$ to $\ket{-1}_e$ and $-1$ to $\ket{0}_e$ since the single-shot readout fidelity is higher for $\ket{-1}_e$. The electron spin is then reset to $\ket{0}_e$ by optical pumping for the next measurement. The parity values of each round are sent to a real-time controller (white block) that determines the feedback operation on the fly, with each round taking approximately 5\,ms. 
% Final measurements 
Since the error correction code protects only against bit-flip errors, we focus on analyzing the population of eigenstates in the $\rm Z$ basis. The final state is examined using three sequential measurements—$\rm ZIZ$, $\rm IZZ$, and $\rm ZII$—of the memory qubits, all read out through the electron spin, along with time-resolved detection of the photonic qubit. Thus, the final state is projected onto one of the sixteen states from $\ket{000}_{c}\ket{e}_{p}$ to $\ket{111}_{c}\ket{l}_{p}$, determined in a single shot using the correlation of the four measurement outcomes. 
Another feature of the final measurement is from the fact that $\rm ZII$ is a logical operator $\rm Z_{L}$. Together with the photonic qubit detection, $\langle\rm Z_{L} Z_{p}\rangle$ can be extracted to determine bit-flip errors at the logical level. At the same time, the measurement outcomes of  $\rm ZIZ$ and $\rm IZZ$ enable error detection of $\langle\rm Z_{L} Z_{p}\rangle$ through post processing.   

We prepare different initial states $\ket{0}_L\ket{e}_{p}$, $\ket{1}_L\ket{l}_{p}$, and  $(\ket{0}_L\ket{e}_{p} + \ket{1}_L\ket{l}_{p})/\sqrt{2}$ , execute the error correction for variable rounds from $\rm M=0$ to $\rm M=12$ with and without real-time feedback. For each experiment setting, we repeatedly run the sequence 20000 times. However, due to low collection and detection efficiency, single photons are detected only in approximately 1\% of experiments. Note the narrow-band frequency filter for selecting ZPL photons is removed for higher data rates, %As a result, phase coherence of the entangled state is expected to be poor. 
while this does not affect the correction of bit-flip errors. 
Part of the results are shown in Figs.~\ref{fig:fig3}\textbf{e-i}, with complete results provided in Appendix.

% Results
Characterizations of the parity measurements are shown in Fig.~\ref{fig:fig3}\textbf{d}, exhibiting expected values for all eight eigenstates with an average readout fidelity of 0.85(1) for $\rm ZIZ$ and 0.84(1) for $\rm IZZ$, excluding fluorescence detection errors. The mapping sequences and error analysis are detailed in Appendix.  % Imperfections in nuclear spin state preparation and parity mapping contribute to comparable errors in these measurements. 
% Repeated stabilizer measurement results
Quantum error correction requires frequent checking and correcting errors before they accumulate beyond the capability of the error correction codes. This process requires non-destructive error-syndrome measurements and fast classical control. We demonstrate this capability by repeatedly performing parity measurements for twelve rounds, though our setup allows even more rounds. 
Using an initial state of $\ket{0}_L\ket{e}_{p}$ as an example, the average values of the two parity checks, with (blue symbols, labeled as ``stabilized") and without (orange symbols, labeled as ``open-loop") active feedback, are plotted as a function of error-correction round index $\rm M$  in Fig.~\ref{fig:fig3}\textbf{e}.  The results show that active feedback substantially slows the decay of both parity checks, indicating the effectiveness in suppressing errors.

% T1 measurement with product states
We extract the joint-state population from the final measurement outcomes and employ an error mitigation method~\cite{shen2012correcting} to correct biases in the average state populations caused by asymmetric errors in state-dependent fluorescence detection (see Appendix). Examples of the final state population distribution are shown in Appendix for three different initial states. Initializing the system in a product state $\ket{0}_L\ket{e}_{p}$ or $\ket{1}_L\ket{l}_{p}$, the fidelity to the initial state can be directly extracted from the final state distribution. In Fig.~\ref{fig:fig3}\textbf{f}, state fidelity to $\ket{0}_L\ket{e}_{p}$ is plotted versus $\rm M$ as blue dots for data with feedback and pink squares for data without feedback. These data points are fitted to the function $p=p_i \exp(-t/T_{1,L})+p_f$, yielding a longer relaxation time $T_{1,L}$ = 31(10) ms and a higher steady population $p_{f}=0.23(7)$ with feedback, compared to $T_{1,L}$ = 20(5) ms and $p_{f}$ = 0.18(4) without feedback. Similar results are observed for $\ket{1}_L\ket{l}_{p}$ as shown in Fig.~\ref{fig:fig3}\textbf{g}, as well as for an entangled state whose total population of $\ket{0}_L\ket{e}_{p}$ and $\ket{1}_L\ket{l}_{p}$ is protected by active feedback (see results in Appendix). 
The observed relaxation of the logical qubit is faster than that of a single memory qubit, limited by state errors introduced by parity mapping operations and parity readout errors both from mapping operations and fluorescence detection. We perform Monte-Carlo simulations and compare the results to experimental data to gain insights into various imperfections for repeated error correction. The comparison for all initial states is shown in Appendix, and we discuss the relevant experimental imperfections and potential ways for improvement in Appendix.

% Error detection results
The final state can be evaluated at the logical level using the measurement of $\langle\rm Z_{L} Z_{p}\rangle$, while the errors in the final state can be detected and rejected in post-processing based on the parity checks of the final measurement. In Fig.~\ref{fig:fig3}\textbf{h}-\textbf{i}, we compare the experimental values of $\langle\rm Z_{L} Z_{p}\rangle$ from all data (squares) to those from the data conditional on ``no error"  (dots), i.e. $\rm ZIZ=+1$ and $\rm IZZ=+1$. In both cases with (Fig.~\ref{fig:fig3}\textbf{h}) and without (Fig.~\ref{fig:fig3}\textbf{i}) feedback,  the results show a clear improvement by error detection. While the improvement is more pronounced when the feedback is not applied, this is because feedback operation accelerates the population accumulation in the  $\ket{1}_{L}\ket{e}_{p}$ and $\ket{0}_{L}\ket{l}_{p}$ states which are $-1$ eigenstates of $\rm Z_{L}Z_{p}$ but $+1$ eigenstates of $\rm ZIZ$ and $\rm IZZ$.

\section{Conclusion and outlook}
In summary, we present coherent and complex control in a hybrid multi-qubit quantum network node. Generated hybrid entanglement enables quantum information transfer between distinct frequency domains. Each of the involved qubits in principle can interact with similar types of qubits in a different system at close frequencies, potentially leading to novel hybrid quantum systems~\cite{kubo2011hybrid,li2015hybrid,li2016hybrid,zhu2014observation,kurizki2015quantum}. Hybrid entanglement can be used for correlated sensing covering a wide frequency range. 
Demonstration of active correction of bit-flip errors can be adapted to correct phase-flip errors and to realize fault-tolerant error correction~\cite{cramer2016repeated,abobeih2022fault}. 
Error correction can be enhanced by improving coherent operations involved in error correction through technical upgrades or quantum optimal control techniques~\cite{xie202399,chou2015optimal},  and by increasing error syndrome readout fidelity using repetitive readout using quantum logic~\cite{jiang2009repetitive}. 
For near-term applications, error detection may be useful to improve the integrity of network performance. 
Our approach can be extended to other platforms, such as trapped ions and atoms, where different type of qubits from the same species~\cite{yang2022realizing,feng2024realization,lai2024realization} or mixed species~\cite{inlek2017multispecies,drmota2023robust} can be used in a similar way as demonstrated in this work. 

Note that during the preparation of this manuscript we became aware of a similar work on hybrid entanglement in a NV center at University of Stuttgart~\cite{javadzade2024efficient}.

\vspace{.5cm}
\noindent\textbf{Acknowledgement} We thank S. Zhang, Y.-F. Pu, Y.-K. Wu, D. Yuan, S. Jiang, and L. Li for helpful discussions.  
This work was supported by the Innovation Program for Quantum Science and Technology (grant nos. 2021ZD0302203, 2021ZD0301601 and 2021ZD0301605), the National Natural Science Foundation of China (grant nos. 12075128, and T2225008), the Tsinghua University Dushi Program, the Shanghai Qi Zhi Institute, the Tsinghua University Initiative Scientific Research Program, and the Ministry of Education of China. L.-M.D. acknowledges in addition support from the New Cornerstone Science Foundation through the New Cornerstone Investigator Program. P.-Y.H. acknowledges the start-up fund from Tsinghua University.

\bibliography{ref}

\end{document}